\documentclass[]{article}
 \usepackage{amsmath}
 \usepackage{graphicx}
 \usepackage{amssymb}
 \usepackage[utf8]{inputenc}
 \usepackage{bigints}
 \usepackage[cal=boondoxo]{mathalfa} 
 \newcommand{\Ai}{{\mathrm{Ai}}}
 \newcommand{\Bi}{{\mathrm{Bi}}}
 \newcommand{\e}{{\mathrm{e}}}
 \numberwithin{equation}{section}
 \everymath{\displaystyle}

\title{Emergent classical universes from initial quantum states in a tomographical description}
\author{C. Stornaiolo \\
	{\em  Istituto Nazionale di Fisica Nucleare,}\\
	{\em Sezione di Napoli,}\\
	{\em  Complesso Universitario di Monte S. Angelo}\\
	{Edificio 6, via Cinthia,45 -- 80126 Napoli, Italy} \\ }

\begin{document}

\maketitle 

\begin{abstract}
	\indent Quantum and classical physical states are represented in a unified way when they are described by symplectic tomography. Therefore this representation allows us to study directly the necessary conditions for   a classical universe to emerge from a quantum state.
	In a previous work  on the de Sitter universe  this was  done by comparing the classical limit of the quantum tomograms with those resulting from the classical cosmological equations. 
	
	In this paper  we first review these results  and  extend them  to all the de Sitter models.  We show further  that these  tomograms can be obtained directly from transposing the Wheeler-De Witt equation to the  tomographic variables. Subsequently, because the classic limits of the quantum tomograms are identified with their asymptotic expressions, we  find the necessary conditions to extend the previous results by taking the tomograms of the WKB solutions of the  Wheeler-DeWitt equation with a any potential.  
	Furthermore in the previous works we found that the de Sitter models undergo the quantum-to-classical transition when the cosmological constant  decays to its present value, we discuss at the end how far we can extend this result to more general models.
	
	In the conclusions,  after discussing  any improvements and developments  of the  results of this work, we sketch a  phenomenological approach from which to extract information about the initial states of the universe.
\end{abstract}

\section{Introduction}
Understanding the universe we live in and its origins is one of the oldest questions of mankind. Only in the last century with the theory of general relativity has it been possible to give a scientific description of the universe and its characteristics. However the study of general relativity has not completely resolved the knowledge of its origins.

The presence of   initial singularity  in the theory, identified with the Big Bang, highlights how the current universe must be the result of a transition from a previous initial state which is generally traced back to an initial quantum state, although, as indicated in \cite{Ellis:2002we}\cite{Ellis:2003qz}\cite{Mulryne:2005ef}, the universe may have emerged from a previous classical state described by a static model and from which the observed characteristics can be derived.

The various stages of the evolution of the universe can be described through the transitions between different physical states, which are not always described in the same way.  

The  state of a physical system is the set of all the informations necessary to determine its evolution, which in turn is  described  a sequence of states. For example in classical physics the state of a particle  is well represented by a point on the phase space, or when the particle is in thermal bath by a distribution function of  probabilities of its position and momentum on a given time.  In classical cosmology, the state of the universe  it should be sufficient to give the expansion factor $ a(t_{0}) $ at some time to determine because it  is a constrained system, but we must  consider also  the  various parameters  present in the in the cosmological equations,  like the spatial curvature $ k $, the present mass-energy density $ \varrho_{0} $, the cosmological constant $ \Lambda $  and so on,  to define the state of the universe as their values determine its evolution.

In Schr\"{o}dinger  picture of quantum mechanics, the state of a particle is given by its wave 
function, which has not a direct physical interpretation, but which is an auxiliary function for defining the expectation values of the various observables quantities. The evolution of a quantum system is  determined by the Schr\"{o}dinger equation. Similarly in quantum cosmology the state of a quantum universe is described by a wave function solution of the Wheeler-DeWitt equation, which  unlike the previous case  does not have an explicit dependence on time. 

The difference of the classical and quantum description does not allow an immediate comparison between these different states and therefore to describe in a simple way the quantum-classical transition of the universe. We can overcome this problem by introducing the symplectic tomographic representation of classical and quantum states which are both  described with the same family of functions\cite{Man'ko:2004zj,Man'ko:2006wv,Capozziello:2007xn,Capozziello:2009ah,Stornaiolo:2014hpa,Stornaiolo:2015goa}.

So to determine  when  a quantum model evolves to a classical universe, it is enough to  check if its  limit, obtained by letting  $ \hbar \to 0$,  converges to the classical tomogram \cite{Stornaiolo:2018lvp,Stornaiolo:2019wau}.  

Another motivation to use the  tomographic representation is  that tomograms  are observables and they allow to determine in laboratory the state of a quantum system (see for example \cite{Raymer:1994},\cite{Kurtsiefer:1997} \cite{Lvovsky:2009zz}).  

Similarly in cosmology, we expect that  the phenomenological determination of the classical tomogram may contain informations of the early quantum state. The possibility of realizing this proposal is based  on Hartle's remarks in \cite{Hartle:1997hw} where he points out two important issues. The first one is that in the case of total absence of informations the density matrix would take the form $ \rho = I/Tr(I) $ where $ I  $ is the unit matrix, but as $ \rho\propto \exp (-H/kT) $, it would correspond to an infinite temperature in equilibrium, which would also correspond to an infinite temperature today, contrarily to the present observations.
The second issue is that even if the entropy observed  $ \tfrac{S}{k}\sim 10^{80} $  is apparently very large, it is however very small compared to the maximum value possible 
$ \tfrac{S}{k}\sim10^{120} $ as showed by Penrose\cite{Penrose}. 

In \cite{Stornaiolo:2018lvp}  \cite{Stornaiolo:2019wau} quantum and classical de Sitter models were analyzed in tomographic  representation. 
The de Sitter models present very simple properties  so that  a complete analytical description was available either in the quantum framework as in the classical.  Therefore their study has allowed to understand the way to face more general models.  In these papers only the Hartle-Hawking model was described in tomographical terms and  it was evident that it does not have a classical limit, however it  was possible to exhibit a tomogram which converged to the classical one derived from  general relativity, Based on more careful considerations, in this work we proof that this  tomogram can be derived from Vilenkin's initial conditions.

Furthermore, a remarkable result  was that in de Sitter's models the quantum-to-classical transition is  induced by the decay of the cosmological constant. To see if this result could be extended to more general models  is the basis of the motivations for this work. However we have gone beyond this problem and we have determined systematic criteria to establish the compatibility between quantum and classical models  starting from asymptotic solutions  of the Wheeler-DeWitt equation with  a generic potential  to represent an extended class of  sources.    

The paper is organized in the following way. In the  first two sections we resume  the general properties for the classical de Sitter models in sect.\ref{sdesitter} and the quantum models in  sect.\ref{squantumdesitter} known from literature. 

In sect.\ref{stomography} we summarize the definitions and  the properties of classical and quantum tomograms.  In sect.\ref{scosmologicaltomography} we take up the results of \cite{Stornaiolo:2018lvp} and complete them with the addition of the Vilenkin and Linde tomogram calculation. 

In sect.\ref{stomographicwheelerdewitt} we complete the work on de Sitter models by introducing a tomographic version of the Wheeler-DeWitt equation and we show that the previous results are obtained as solutions of this equation.

In sect.\ref{sscomparisontomograms} we show that taking the classical limit $ \hbar \to 0 $  of the de Sitter quantum tomograms is equivalent to take the asymptotic expressions of the tomograms and  first we  check if they converge  to (\ref{classicaldeSitterBoltzmann})  the classical tomogram and then we observe that the same asymptotic limits are obtained taking the limit $ \lambda\to\lambda_{today} $ implying that the classical state of the universe may be the result of the decay of the cosmological constant.

We extend these results  to a more general setting. In sect.\ref{swkb} we look for a general definition of the classical limit of the quantum tomograms and  by  considering the WKB approximation in the Wheeler-DeWitt equation with a generic potential $ V(q) $ and then we calculate their tomograms. This allows us  to determine which quantum tomograms converge to the classical ones and which do not. In sect.\ref{smodelswithcosmologicalconstant} we consider  general models with  the cosmological constant.

In sect.\ref{sConclusions} we discuss the results, any improvement and the  perspectives of this work.

\section{The de Sitter models}\label{sdesitter}

Albert Einstein introduced the cosmological constant  in order to have  a static universe as it was expected at the beginning of the XXth century, while  his original  theory predicted a dynamical universe. But  in 1917 Willem de Sitter showed that the static model was unstable and  that the cosmological constant played instead  an important dynamical role even in absence of material sources.

The dynamics of the de Sitter  is very simple and describes an  everlasting exponential expanding universe with an initial singularity in infinity past.

The de Sitter model has taken on an important role to introduce  the inflationary paradigm needed   to explain first of all in a generic way  the flatness and homogeneity  of the universe without recurring to a fine tuning of the initial conditions.  The role of cosmological constant was attributed to a false vacuum state of the inflaton (a hypothetical particle responsible for this stage of the universe),  represented by a stationary state (minimum or maximum) in the potential of this field.

Presently  the cosmological constant is considered together with other dark matter  candidates to explain  the accelerated  expansion of the universe as  the observations on Type 1A  supernovae indicates.

The physical interpretation of the cosmological constant is  related by many authors  to  the quantum vacuum energy.  But this interpretation  poses a serious  problem on why the cosmological constant is so small compared with the vacuum energy expected. In his seminal paper \cite{Weinberg:1988cp} S. Weinberg estimates the value of the vacuum energy  which  should be larger than the actual value and eventually proportional to the Planck energy density where its estimated value of $ \Lambda $ is  $ 10^{122} $ orders of degree smaller. 

In this paper we describe the de Sitter  spacetime with the metric \cite{Louko:1987wq}

\begin{equation}\label{metric}
ds^{2}=-\frac{N^2}{q}d\tau^{2}+\frac{q}{1-kR^{2}}dr^{2}+qr^{2}d\Omega_{2}
\end{equation}
where respect to the conventional metrics  $ q=a^{2} $ (where $ a $ is the expansion factor)  and  the lapse is  scaled by a factor $ \sqrt{q} $.  In the following only the case $ k=1 $ will be considered.  

From the Einstein equations with cosmological constant
\begin{equation}\label{vacuumEinstein equations}
G_{ab}+\Lambda g_{ab}=8\pi G T_{ab}
\end{equation}
and with $ N=N(t)$,  we have the following equations

\begin{equation}\label{equationsinloukoform1}
\frac{1}{4}\frac{1}{N^{2}}\frac{\dot{q}^{2}}{q}+\frac{1}{q}=\frac{8\pi G}{3}T_{00}+ \frac{\Lambda}{3}
\end{equation}
and 

\begin{equation}\label{equationsinloukoform2}
\frac{\ddot{q}}{N^{2}}+\frac{1}{2} \frac{1}{N^{2}}\frac{\dot{q}^2}{q}-\frac{\dot{N}}{N^{3}}\dot{q} +\frac{2}{q}= -\frac{8\pi G}{3}T+\frac{4}{3}\Lambda\, .
\end{equation}
The previous  equations (putting for sake of simplicity $ N=1 $) imply the conservation equation 
\begin{equation}\label{conservationequation}
\dot{T}_{00}=\frac{1}{2}\frac{\dot{q}}{q}\left(T-4T_{00} \right) \, .
\end{equation}
de Sitter's model contemplates only the cosmological constant for which we will pose $ T_{ab}=0 $ in the rest of the section.
We notice that in the gauge  $N=\sqrt{q}  $ we obtain the classical solution
\begin{equation}\label{usualclassicalsolution}
q(t)=\Lambda^{-1}\cosh^{2}(\Lambda^{1/2}t)\, .
\end{equation}
For numerical reasons,  in the following we  multiply the metric by a factor $ \sigma^{2}=2G/3\pi $.

In order to quantize the gravitational system we need move on to  the  Hamiltonian formalism.  
First we write the action, which after integrating the spatial part and eliminating a total derivative with respect to time,   is \cite{Louko:1987wq}
\begin{equation}\label{action2}
S=\frac{1}{2}\int  N\left( - \frac{ \dot{q}^{2}}{4  N   } +1-\lambda q\right)  dt
\end{equation}
where $ \lambda $ is now the cosmological constant  in Planck units. 
The coordinates of the phase space are $ (q,p) $  with  the momentum  $ p  $ defined by
\begin{equation}\label{pmomentum}
p=\frac{\partial L}{\partial \dot{}q}=-\frac{\dot{q}}{4N}
\end{equation}
using the Legendre transform,  the action takes the form
\begin{equation}\label{action3withhamiltonian}
S=\int \left( p\dot{q}-N\mathcal{H} \right) dt
\end{equation}
with
\begin{equation}\label{HamiltoniandeSitter}
\mathcal{H}=\frac{1}{2}\left( -4p^{2}+\lambda q-1\right) 
\end{equation}
The lapse function  $ N $ is  a Lagrange multiplier,   the variation 
\begin{equation}\label{variationwithrespectto N}
\frac{\delta S}{\delta N }=0 
\end{equation}
implies  the Hamiltonian constraint,
\begin{equation}\label{constraint}
\frac{1}{2}\left( -4p^{2}+\lambda q-1\right) =0
\end{equation}
which  is equivalent to equation (\ref{equationsinloukoform1}). 

The state of a  classical  system at a time $ t _{0}$ is given by the values of the   positions and  the   momenta  at that  time.  For a constrained system this values are restricted to a submanifold of the phase space. Eq. (\ref{constraint}) describes  the phase space a curve of the  states that the universe is allowed to take. 

To express the state of the universe  as a distribution on the phase space, let us  assume that the universe is composed by $ N $ local subsystems with $ N $ is arbitrarily  large, all with the same metric (\ref{metric}), but  following a statistical behavior that can be described  by  Boltzmann equation
\begin{equation}\label{classicaldeSitterBoltzmann}
\frac{\partial f(q,p)}{\partial t} + \left\lbrace f,\mathcal{H} \right\rbrace  =0\, ,
\end{equation}
where $ \left\lbrace \right\rbrace  $ are the Poisson brackets, but if
\begin{equation}\label{zerocondition}
\frac{\partial f(q,p)}{\partial t}=0.
\end{equation}
the   solutions  to the Boltzmann equation are    the functions $ f(q,p) $ which commute with $ \mathcal{H} $,  the most simple are just of the form $ f(\mathcal{H}) $.

One choice can be 
\begin{equation}\label{classicaldesitterdistribution}
f(q,p)=\delta\left( -4p^{2}+\lambda q-1\right) .
\end{equation}
which is equivalent to say that there is  a strict homogeneity of all the subsystems composing the universe. Another choice could be
\begin{equation}\label{gaussianboltzmann}
f_{gauss}(q,p)=\exp\left[-\left( -4p^{2}+\lambda q-1\right)^{2} \right] 
\end{equation}
in which there is a Gaussian fluctuation of the metric. Indeed it describes a universe where there are subsystems deviating  from the   condition  (\ref{constraint}).  Such deviations are equivalent to local variations of the spatial curvature $ k $. Because when   $ -4p^{2}+\lambda q-1\neq 0 $  it must be
take a value $ x_{0} $ and then $$ -4p^{2}+\lambda q-k=0 $$
and where  $ k=1+x_{0} $. 
In   the following we shall work only with    (\ref{classicaldesitterdistribution}).

\section{Quantizing the de Sitter model and the initial conditions}\label{squantumdesitter}
In quantum cosmology the states of  the universe are  described by the wave functions $ \Psi(q) $ defined on the minisuperspace, i.e. the space of the homogeneous metrics.   They are solutions  of the Wheeler-DeWitt equation, which is  obtained by substituting the Hamiltonian constraint with the equation
\begin{equation}\label{hamiltonian operator}
\hat{\mathcal{H}}\Psi(q)=0
\end{equation}
where $ \hat{\mathcal{H}} $ is the  Hamiltonian operator obtained by substituting $ q $ and $ p $ in (\ref{constraint}) with the respective operators
\begin{equation}\label{operator p}
\mathbf{\hat{p}}\Psi=-i\hbar\frac{d\Psi}{dq}
\end{equation}
and
\begin{equation}\label{operator q}
\mathbf{\hat{q}}\Psi(q)=q \Psi(q)\, .
\end{equation}

The Wheeler-DeWitt equation corresponding to the  Hamiltonian (\ref{HamiltoniandeSitter}) is \cite{Louko:1987wq}
\begin{equation}\label{WheelerDeWitt}
\left(4\hbar^{2}\frac{d^{2}}{dq^{2}}+\lambda q -1 \right)\psi(q) =0\, .
\end{equation}
It  can be reduced to the Airy equation   \cite{abramowitz,Vallee}
\begin{equation}\label{Airyequation}
\frac{d^{2}\psi(\xi)}{d \xi^{2}}-\xi\psi(\xi)=0,  
\end{equation}
with the  change of variable 
\begin{equation}\label{changeofvariable1}
\xi =\frac{1-\lambda q}{\left(2\hbar\lambda \right)^{2/3} }\, .
\end{equation}

This equation has  two independent solutions $ \Ai(x) $ and $ \Bi(x) $ whose   integral representations are respectively  {\cite{abramowitz}, \cite{Vallee} }
\begin{equation}\label{AiryAi}
\Ai(x)=\dfrac{1}{2\pi}\int_{-\infty}^{+\infty}\exp\left[ {i\left( \dfrac{z^{3}}{3}+xz\right) }\right] dz
\end{equation}
and 
\begin{equation}\label{AiryBi}
\Bi(x)=\dfrac{1}{\pi}\int_{0}^{+\infty}\left[ \exp   \left( -\dfrac{z^{3}}{3}+xz\right) +\sin\left( \dfrac{z^{3}}{3}+xz\right) \right] dz\, .
\end{equation}
All the solutions of the Airy equation are  linear combinations  of  $ \Ai(x) $ and $ Bi(x) $.  

The different proposals given by  Hartle and Hawking,   Vilenkin and   Linde's  are characterized by different combinations of these functions (see for example in  \cite{Louko:1987wq}, \cite{Halliwell:1988ik},  \cite{Cordero:2011xa}).

Indeed the Hartle and Hawking "No Boundary Condition" corresponds to the  wave function \cite{Hartle:1983ai}   
\begin{equation}\label{HartleHawkingwavefunction}
\psi_{HH}=\mathrm{a}\, \Ai\left( \frac{1-\lambda q}{\left(2\hbar\lambda \right)^{2/3} }\right), 
\end{equation} 
Vilenkin's  "Tunneling from Nothing" \cite{Vilenkin:1986cy} \cite{Vilenkin:1987kf} is  represented by the combination 
\begin{equation}
\label{Vilenkinwavefunction}
\psi_{V}\left( \frac{1-\lambda q}{\left(2\hbar\lambda \right)^{2/3} }\right) =\frac{\mathrm{b}}{2}\left( \Ai\left( \frac{1-\lambda q}{\left(2\hbar\lambda \right)^{2/3} }\right) +i\, \Bi\left( \frac{1-\lambda q}{\left(2\hbar\lambda \right)^{2/3} }\right) \right)
\end{equation}
and finally the initial condition discussed by Linde which is an extension of Vilenkin's proposal is  \cite{Linde:1983mx} is 
\begin{equation}
\label{Lindewavefunction}
\psi_{L}=-i \mathrm{c} \,\Bi\left( \frac{1-\lambda q}{\left(2\hbar\lambda \right)^{2/3} }\right)
\end{equation}
with  a ,  b  and  c  are normalization constants.  

When $ q>1/\lambda $ the argument of   the Airy functions is  negative  and these functions oscillate,   while for $q<1/\lambda$,    $ \Ai(\tfrac{1-\lambda q}{(2\hbar\lambda)^{2/3}}) $ decays rapidly  and $ \Bi(\tfrac{1-\lambda q}{(2\hbar\lambda)^{2/3}}) $ grows to infinity.
These functions are better characterized by their asymptotic behavior when  $ q>1/\lambda $. 

In fact for  Vilenkin's  wave function becomes  
\begin{equation}\label{psivil}
\psi_{V}\approx \mathrm{b}\frac{(2\pi \hbar)^{1/6}}{2[\pi^{2}(\lambda q-1)]^{1/4}}\exp(-i\xi)\, , 
\end{equation}
the Hartle and Hawking wave function takes the form 
\begin{equation}\label{psihar}
\psi_{HH}\approx \mathrm{a}\frac{(\pi \hbar)^{1/6}}{[\pi^{2}(\lambda q-1)]^{1/4}}\cos\left(  \xi-\frac{\pi}{4}\right) 
\end{equation}
and Linde's wave function takes
\begin{equation}\label{psilind}
\psi_{L}\approx i\mathrm{c}\frac{(\pi \hbar)^{1/6}}{[\pi^{2}(\lambda q-1)]^{1/4}}\sin\left(  \xi-\frac{\pi}{4}\right) \, .
\end{equation}
where  $ \xi= \tfrac{2}{3}(\lambda q-1)^{3/2}$). These three expressions are derived  either  formallyfrom their analytic definitions (\ref{AiryAi}) and (\ref{AiryBi}) (see \cite{abramowitz} and\cite{Vallee}), or by   applying the WKB approximation to eq.(\ref{Airyequation}) (see \cite{Zwillinger}).  They show  that Vilenkin's wave function (\ref{psivil}) presents only expanding modes, whereas the other two wave functions present  combinations of expanding and collapsing modes
(see also the discussion in sect. \ref{swkb}).

\section{Classical and quantum states in tomographic  representation}\label{stomography}

The aim of symplectic tomography is to reconstruct a physical state (classical or quantum) by using a set of marginal functions. A classical  physical state is represented on the phase space by a distribution function $ f(q,p) $  solution of the Boltzmann equation.  A quantum state is described by a wave function or by a  Wigner  function $ W(q,p)$ on the phase space.  The Wigner function is called a quasi-distribution because it  can take negative values, whereas  a classical distribution function is always positive.

In order  to represent  quantum and classical state in a homogeneous way we can  describe them  in terms of symplectic tomography. For an introduction on this topic  see for example \cite{Manko:1999ydp} and \cite{Ibort:2009bk}.

Let us consider a distribution on a one dimensional  phase space where al the states are represented by a pair of coordinates $( q,p )$.  We consider the projection  of this distribution  on the $ q $ axis i.e we consider a  function $ W(q) $.  If we rotate the coordinate system   by the transformations 
\begin{equation}\label{X}
X=\mu q+\nu p
\end{equation}
and 
\begin{equation}\label{P}
P=-\nu q+\mu p
\end{equation}
with  $ \mu=s \cos \theta $ and $ \nu= s^{-1}\sin \theta  $, where $ s $ is a squeezing factor  and $ \theta $ is the rotation angle of the $ (q, p) $ frame, we can take all the  projections $ \mathcal{W}(X,\mu,\nu) $   of the same distribution on each $ X $ axis (where  the initial $ W(q)=\mathcal{W}(X,1,0)$).  The set of all these projections is the tomogram. So we can reconstruct the distribution by appropriately treating the tomogram. This is done generally with an inverse Radon transform as we will see in the following. Other reconstructing algorithms  are  used when a tomogram is obtained from a set of experimental data   \cite{Lvovsky:2009zz}. For this reason determining the  tomogram of a system is equivalent to determine its state. It is important to remark that these transformations are  linear canonical transformations\cite{Manko:1999ydp}.
Let us define the  classical and quantum tomogram.

Given a  classical probability  distribution $ f(q,p) $ on the phase space, we define  the  classical  tomogram  by
\begin{equation}\label{tomoboltz}
\mathcal{W}\left(X,\mu,\nu \right)=
\int f(q,p)\delta(X-\mu q-\nu p)dq dp\,.
\end{equation}
If  $ f(q,p) $ is normalized then also the tomogram is normalized and satisfies the following conditions,
\begin{equation}\label{normalization condition tomogram}
\int \mathcal{W}\left(X,\mu,\nu \right) dX=1
\end{equation}
\begin{equation}\label{tomogramposition}
\mathcal{W}\left(X,1,0 \right) =\int f(q,p)dp
\end{equation}
and
\begin{equation}\label{tomogrammomentum}
\mathcal{W}\left(X,0,1 \right)=\int f(q,p) dq\,.
\end{equation}
Similarly  a quantum tomogram defined by
\begin{equation}\label{tomowigner}
\mathcal{W}\left(X,\mu,\nu \right)=
\int W(q,p)\delta(X-\mu q-\nu p)dq dp\,. 
\end{equation}
where $ W(q,p) $ is the Wigner function. This is a general definition for pure and mixed states.  For a pure state, given a  wave function $ \psi(q) $,  an equivalent definition of tomogram is
\begin{equation} \label{nunonzero}
\mathcal{W}\left(X,\mu,\nu \right)=
\frac{1}{2\pi\hbar|\nu|}\left| \int \psi\left( y\right)  \exp\left[ {i\left( \frac{\mu}{2\hbar\nu}y^{2}-\frac{X}{\hbar\nu}y\right)}\right] dy\right| ^{2}\, .
\end{equation}
In particular if  $ \nu=0 $  and $ \mu=1 $ we have
$$ 	\mathcal{W}\left(X,1,0 \right)=\frac{1}{2\pi}\left|  \psi\left(X\right)  \right| ^{2}\equiv \frac{1}{2\pi}\left|  \psi\left(x\right)  \right| ^{2} .$$
and for  $ \mu=0 $  and $ \nu=1 $  we have the  Fourier transform with respect to the variable $ p/\hbar $,  

\begin{equation} \label{mu0}
\mathcal{W}\left(X,0,1 \right)=\frac{1}{2\pi\hbar}
\left| \int \psi\left( y\right)  \exp\left[ {-i\frac{ py}{\hbar}}\right]   dy\right|^{2}\,.
\end{equation}
i.e. it   is the square modulus of the wave function in the $ p $ representation. 

The symplectic tomograms $\mathcal{ W}(X,\mu, \nu)  $ are probability functions and  satisfy the following conditions
\noindent
1) Nonnegativity
\begin{equation}\label{nonnegativity}
\mathcal{W}(X,\mu,\nu)\geq 0\, ,
\end{equation}
\noindent
2) Normalization
\begin{equation}\label{normalization}
\int\mathcal{ W}(X,\mu, \nu) dX = 1\, , 
\end{equation}
and 
\noindent
3)  homogeneity
\begin{equation}\label{homogeneity}
\mathcal{ W}(\alpha X,\alpha\mu, \lambda\nu) = \frac{1}{|\alpha|}
\mathcal{ W}(X,\mu, \nu) . 
\end{equation}
The important thing is that all these relations can be inverted \cite{Capozziello:2007xn}\cite{Stornaiolo:2018lvp}\cite{Manko:1999ydp}\cite{Ibort:2009bk}\cite{Vogel:1989zz} and then tomograms represent the state of a classical or a quantum system. They are observables and  have been used to reconstruct the Wigner function  in quantum mechanics and quantum optics   \cite{Vogel:1989zz}\cite{Janicke:1989}\cite{Raymer:1994}\cite{Kurtsiefer:1997}\cite{Lvovsky:2009zz}.

Quantum and classical tomograms differ in that the former must satisfy the uncertainty principle in the form stated by Robertson\cite{Robertson} and Schr\"{o}dinger\cite{Schroedinger} (see also \cite{dodonovmanko1}-\cite{dodonov2}), 
which is  expressed in terms of the variance of the canonical variables $ \sigma_{pp}$,  $ \sigma_{qq}$ and the covariance $ \sigma_{qp}  $,  
\begin{equation}\label{RobertsSchroedinger}
\sigma_{qq}\sigma_{pp}-\sigma_{qp}^{2}\geq \frac{1}{4}\hbar^{2}\, .
\end{equation}
Unlike the uncertainty principle enunciated by Heisenberg, it  
is invariant under the group of linear canonical transformations\cite{sudarshan} so  in a tomogram if it is  verified  for a pair $ (\mu, \nu) $,  it is true for every other pair.

However in many case it is sufficient to write the uncertainty principle in its weakest form ( with $ \sigma_{qp}=0 $),

\begin{align}\label{tomographic uncertainty relation}
\left[\int  \mathcal{ W}(X,1, 0)X^{2}dX\right. &-\left. \left\lbrace\int \mathcal{ W}(X,1, 0)XdX \right\rbrace^{2} \right]\\
&  \times\left[\int  \mathcal{ W}(X,0, 1)X^{2}dX-\left\lbrace\int \mathcal{ W}(X,0, 1)XdX \right\rbrace^{2} \right] \geq \frac{1}{4}\hbar^{2}
\end{align}
If this  last  condition is violated certainly the tomogram is related  to a classical system.

\section{Tomographic description of the de Sitter universe}\label{scosmologicaltomography}
Let us now analyze the classical and quantum  de Sitter models from the  tomographic point of view. 

If we  apply (\ref{tomoboltz}) to  the distribution  (\ref{classicaldesitterdistribution})
\begin{equation}\label{classicaluniverseconstraint}
f(q,p)=\delta\left( -4p^{2}+\lambda q-1\right) .
\end{equation}
we find the  classical de Sitter tomogram
\begin{equation}\label{classicaluniversedetomogram}
\mathcal{W}\left(X,\mu,\nu \right)=\frac{1}{2|\mu|}\frac{1}{\left| \sqrt{\frac{\lambda^{2}\nu^{2}}{16\mu^{2}}+\frac{\lambda X}{\mu}-1}\right| }\, .
\end{equation}
It is clear that this  function  is not  integrable on an infinite range,  then its domain must be restricted to  compact support such that the normalization condition (\ref{normalization condition tomogram}) is satisfied.   We    choose a  closed interval $ I $ such that  the  normalization condition is satisfied. For example if we  choose   the inferior value of the interval is such that
\begin{equation} 
\frac{\lambda^{2}\nu^{2}}{16\mu^{2}}+ \frac{\lambda X}{\mu}-1= 0
\  \  \  \ {\rm i.e.} \ \ \ 
X= \frac{\mu}{\lambda} \left( 1-\frac{\lambda^{2}\nu^{2}}{16\mu^{2}}\right) 
\end{equation}
we see that 
\begin{equation}
\int_{-\infty}^{\infty} \mathcal{W}(X,\mu,\nu)dX= \frac{1}{2|\mu|}\int_{\frac{\mu}{\lambda} \left( 1-\frac{\lambda^{2}\nu^{2}}{16\mu^{2}}\right) }^{\lambda\mu+\frac{\mu}{\lambda} \left( 1-\frac{\lambda^{2}\nu^{2}}{16\mu^{2}}\right) }\frac{dX}{\left| \sqrt{\frac{\lambda^{2}\nu^{2}}{16\mu^{2}}+\frac{\lambda X}{\mu}-1}\right| }=\frac{\mu}{\left|\mu \right| }=1\, .
\end{equation}
if $ \mu>0 $.

In conclusion    the value of the cosmological constant  of the order of  $ 10^{-122}\lambda_{Planck} $,  implies that  the interval $ I $ becomes very  narrow and  the classical tomogram of a de Sitter  universe  is approximated by a  delta function, 

\begin{equation}
\mathcal{W}\left( X,\mu,\nu\right) \approx \delta\left( \frac{\lambda}{\mu}X +\frac{\lambda^{2}\nu^{2}}{16 \mu^{2}}-1\right) \, ,
\end{equation}
which means  that the extreme smallness of the cosmological constant implies a  very high degree of  homogeneity of the classical de Sitter universe. To loose the restrictions imposed by the cosmological constant we can define the tomogram on   a larger interval $ [X_{1},X_{2}] $  provided to normalize the tomogram  by 
\begin{equation}\label{normalizedtomogram}
\bar{\mathcal{W}}\left( X,\mu,\nu\right)=\dfrac{\mathcal{W}\left( X,\mu,\nu\right)}{\int_{X_{1}}^{X_{2}}\mathcal{W}\left( X,\mu,\nu\right)}
\end{equation}

Now let's determine the quantum tomograms from the wave functions (\ref{HartleHawkingwavefunction})--(\ref{Lindewavefunction})
by  applying the first  fractional Fourier transform (\ref{nunonzero}) and  taking its square modulus of   and divided by $2\pi\hbar |\nu| $.  
For  (\ref{HartleHawkingwavefunction})  we have,  
\begin{equation}\label{fractionalfourierhartlehawking}
\Psi_{HH}(X,\mu,\nu)=A\int \exp\left( i\left( \frac{z^{3}}{3}+ \frac{1-\lambda q}{(2\hbar \lambda)^{2/3}}z+\frac{\mu}{2\hbar\nu}q^{2}-\frac{Xq}{\hbar\nu}\right) \right) dqdz
\end{equation}
and the Hartle-Hawking tomogram is 

\begin{equation}\label{tomo_hh}
\mathcal{W}_{HH}(X,\mu,\nu)= \frac{\mathrm{a}^{2}}{2\pi\hbar|\mu|}\left|\Ai\left( \frac{1}{(2\hbar\lambda)^{2/3}}\left( 1-\frac{\lambda X}{\mu}-\frac{\lambda^{2}}{16}\frac{\nu^{2}}{\mu^{2}}\right) \right)  \right|^{2} \, .
\end{equation}
It is straightforward to    extend this calculation to the Vilenkin and Linde wave functions by noticing that ,  the  wave functions  (\ref{Vilenkinwavefunction}) and (\ref{Lindewavefunction} )   can be  written as  \cite{abramowitz}  \cite{Vallee}
\begin{equation}
\label{Vilenkin}
\psi_{V}\left( \frac{1-\lambda q}{\left(2\hbar\lambda \right)^{2/3} }\right)
=\mathrm{b}e^{i\pi/3}\Ai\left(e^{-2\pi i/3}  \frac{1-\lambda q}{\left(2\hbar\lambda \right)^{2/3} }  \right)
\end{equation}
and  
\begin{equation}
\label{Linde}
\psi_{L}\left( \frac{1-\lambda q}{\left(2\hbar\lambda \right)^{2/3} }\right)=\mathrm{c}\left[ e^{4\pi i/3}\Ai\left(e^{4\pi i/3}  \frac{1-\lambda q}{\left(2\hbar\lambda \right)^{2/3} }\right)+
e^{2\pi i/3}\Ai\left(e^{2\pi i/3} \frac{1-\lambda q}{\left(2\hbar\lambda \right)^{2/3} }  \right)\right]\, .
\end{equation}
and finally we find that the Vilenkin and Linde tomograms are respectively  
\begin{align}\label{tomo_v}
\nonumber \mathcal{W}_{V}(X,\mu,\nu)&=\frac{\mathrm{b}^{2}}{2\pi\hbar|\mu|}\left|\Ai\left( \frac{1}{(2\hbar\lambda)^{2/3}}\left( 1-\frac{\lambda X}{\mu}-\frac{\lambda^{2}}{16}\frac{\nu^{2}}{\mu^{2}}\right) \right)\right. \\
&\left. +i\Bi\left( \frac{1}{(2\hbar\lambda)^{2/3}}\left( 1-\frac{\lambda X}{\mu}-\frac{\lambda^{2}}{16}\frac{\nu^{2}}{\mu^{2}}\right) \right)  \right|^{2} 
\end{align}
and
\begin{equation}\label{tomo_l}
\mathcal{W}_{L}(X,\mu,\nu)=\frac{\mathrm{c}^{2}}{2\pi\hbar|\mu|}\left|\Bi\left( \frac{1}{(2\hbar\lambda)^{2/3}}\left( 1-\frac{\lambda X}{\mu}-\frac{\lambda^{2}}{16}\frac{\nu^{2}}{\mu^{2}}\right) \right)  \right|^{2} 
\end{equation}

\section{The Wheeler-DeWitt tomographic equation}\label{stomographicwheelerdewitt}

In this section we show that can find the tomograms (\ref{tomo_hh}), (\ref{tomo_v})  and (\ref{tomo_l})   as solutions  of  the Wheeler-DeWitt equation expressed with the  tomographical variables $ X $, $ \mu $ and  $ \nu $.

This equation is obtained by introducing  the  correspondences  between operators,
\begin{equation}\label{q to tomographic variables}
q\to -\left( \frac{\partial}{\partial X}\right) ^{-1} \frac{\partial}{\partial\mu}+i\frac{\nu}{2\hbar}\frac{\partial}{\partial X}
\end{equation}
and 
\begin{equation}\label{dsu dq to tomographicvariables}
\frac{d}{dq}\to \frac{\mu}{2}\frac{\partial}{\partial X}-i\hbar\left( \frac{\partial}{\partial X}\right) ^{-1}\frac{\partial}{\partial\nu}
\end{equation}
which applied to (\ref{WheelerDeWitt})
gives the tomographic Wheeler-DeWitt equation
\begin{align}\label{tomwdw}
\left(  \hbar^{2}\mu^{2}\frac{\partial^{2}}{\partial X^{2}}\right. &-4i\hbar\mu\frac{\partial}{\partial \nu} -4\left( \frac{\partial}{\partial X}\right) ^{-2}
\frac{\partial^{2}}{\partial \nu^{2}}\\
&  \left.  -\lambda\left( \frac{\partial}{\partial X}\right) ^{-1}\frac{\partial}{\partial\mu}+i\frac{\hbar\lambda\nu}{2}\frac{\partial}{\partial X}-1\right) \Psi(X,\mu,\nu)=0
\end{align}
A tomogram is obtained  by the  square modulus of one of the solutions of this equation.

From the imaginary part we obtain the equation
\begin{equation}\label{imaginary}
-4\mu\frac{\partial\Psi}{\partial \nu} +\frac{\lambda\nu}{2}\frac{\partial\Psi}{\partial X}=0
\end{equation}
that   gives us the expression
\begin{equation}\label{imaginaryelaborated}
\frac{\partial^{2}\Psi}{\partial\nu^{2}}=
\frac{\lambda}{8\mu}\frac{\partial\Psi}{\partial X}
+\left( \frac{\lambda\nu}{8\mu}\right)^{2}\frac{\partial^{2}\Psi}{\partial X^{2}}
\end{equation}
that can be inserted in the real part to obtain finally
\begin{equation}\label{realtomwdw}
\hbar^{2}\mu^{2}\frac{\partial^{2}\Psi}{\partial X^{2}}
-\frac{\lambda}{2\mu}\left( \frac{\partial}{\partial X}\right) ^{-1}\Psi
-\left(\frac{\lambda\nu}{4\mu} \right)^{2}\Psi
-\lambda\left( \frac{\partial}{\partial X}\right) ^{-1}\frac{\partial\Psi}{\partial\mu}-\Psi=0
\end{equation}
To solve this equation we first take the Fourier transform of $ \Psi $, $ \tilde{\Psi}(k,\mu,\nu) $ so that equation (\ref{realtomwdw}) becomes
\begin{equation}\label{fouriertransform}
\frac{\partial \tilde{\Psi}}{\partial \mu}=\left( -i\left( \frac{\hbar^{2}\mu^{2}k^{3}}{\lambda}- \lambda\left(\frac{\nu}{4\mu} \right)^{2}k -\frac{k}{\lambda}\right)  -\frac{1}{2\mu }\right) \tilde{\Psi}
\end{equation}
which integrated gives
\begin{equation}\label{fouriertransformsolution}
\tilde{\Psi}(k,\mu,\nu)=\frac{\mathrm{A}}{\sqrt{\mu}} \exp\left[ -i\left(\frac{\hbar^{2}\mu^{3}k^{3}}{3\lambda}+\left( \frac{\mu}{\lambda}-\frac{\lambda\nu^{2}}{16\mu}\right)k  \right) \right] 
\end{equation}
and we finally we obtain
\begin{equation}\label{fourierantitransform}
\Psi(X,\mu,\nu)=\frac{\mathrm{A}}{\sqrt{\mu}}\int_{-\infty}^{\infty}\exp\left[ -i\frac{\hbar^{2}\mu^{3}k^{3}}{\lambda}-i\left( \frac{\mu}{\lambda}-\frac{\lambda\nu^{2}}{16\mu}-X\right)k   \right]dk
\end{equation}
using the relation
\begin{equation}\label{airyformula}
\Ai(ax)=\frac{1}{2\pi a}\int_{-\infty}^{\infty}e^{i\frac{z^{3}}{3a^{3}}+ixz}dz
\end{equation}
we obtain the following solutions of the Airy equation  
\begin{equation}\label{partialfourierpsi}
\Psi(X,\mu,\nu)=\mathrm{A}\frac{e^{in\frac{\pi}{3}}\mu^{1/2}\hbar^{2/3}}{\lambda^{1/3}}\Ai\left( \frac{e^{i(1-\frac{n}{3})\pi}}{(\hbar\lambda)^{2/3}}\left( 1-\frac{\lambda X}{\mu}-\frac{1}{16}\frac{\nu^{2}}{\mu^{2}}\lambda^{2}\right) \right).
\end{equation}
where $ A $ is an arbitrary integration coefficient which to be determined by requiring that  the asymptotic approximations have the same coefficient of  the classical tomogram.  

Finally the tomogram is given by 
\begin{equation}\label{prefinal tomogram}
W(X,\mu,\nu)=\left| \Psi(X,\mu,\nu)\right| ^{2}
\end{equation}

Which becomes, after applying the scaling property of the tomograms,
\begin{equation}\label{scaling}
W(\alpha X, \alpha \mu, \alpha \nu)=\frac{1}{|\alpha|}W(X,\mu,\nu)
\end{equation}
\begin{equation}\label{final tomogram}
W(X,\mu,\nu)=\mathrm{A}^{2}\frac{\hbar^{4/3}}{|\mu|\lambda^{2/3}}\left|\Ai\left( \frac{e^{i(1-\frac{n}{3})\pi}}{(\hbar\lambda)^{2/3}}\left( 1-\frac{\lambda X}{\mu}-\frac{1}{16}\frac{\nu^{2}}{\mu^{2}}\lambda^{2}\right) \right) \right|^{2}
\end{equation}

The values  $ n=3 $ and  $ n=1 $ correspond respectively  to the  Hartle-Hawking  and   to  the Vilenkin tomograms. The Linde solution is a linear combination of two independent solutions of eq.  (\ref{tomwdw}). 

We notice that there is a slight difference with the functions obtained with the fractional Fourier transforms (\ref{tomo_hh}), (\ref{tomo_v})  and (\ref{tomo_l}) by a factor $ 2^{2/3} $ in the denominator of the argument, this difference can be remedied by changing the definition of the operators (\ref{q to tomographic variables}) and (\ref{dsu dq to tomographicvariables}).

\section{Classical limits of the de Sitter tomograms and  the classical tomogram}\label{sscomparisontomograms}

In this section discuss  the relation between the quantum and classical tomograms.  To this aim we look at  the classical limit of a quantum tomogram  ($ \hbar \to  0 $), but for  the  tomograms (\ref{tomo_hh}), (\ref{tomo_v})  and (\ref{tomo_l})
we can take  the limit  $  (2\hbar\lambda)^{2/3}\to 0 $, which leads to  their asymptotic expansions. They  coincide with the asymptotic expansions of the Airy functions which  change depending on whether  argument is positive or negative. In the first case we have an exponential growth, for the Bi function and an exponential decrease for the Ai function. In the second case both functions oscillate. 

Let us call the argument of the de Sitter tomograms $$ S=\frac{1}{3\hbar\lambda}\left( 1-\frac{\lambda X}{\mu}-\frac{\lambda^{2}}{16}\frac{\nu^{2}}{\mu^{2}}\right)^{3/2} -\frac{\pi}{4}.  $$

For   $  S<0 $
then  the Hartle and Hawking tomogram 
is asymptotically
\begin{align}\label{negativeAi}
\mathcal{W}_{HH}(X,\mu,\nu)&\approx   \frac{\mathrm{a}^2}{8\pi^{2}\hbar|\mu|}\frac{(2\hbar\lambda)^{1/3}}{\left|1-\frac{\lambda X}{\mu}-\frac{\lambda^{2}}{16}\frac{\nu^{2}}{\mu^{2}}\right|^{1/2}}
\times \left|   \cos\left(  S  \right)  \right|^{2} , \\  
\end{align}
when  $S>0 $
\begin{equation}\label{positiveAi}
\mathcal{W}_{HH}(X,\mu,\nu)\approx   \frac{\mathrm{a}^2}{16\pi^{2}\hbar|\mu|}\frac{(2\hbar\lambda)^{1/3}}{\left|1-\frac{\lambda X}{\mu}-\frac{\lambda^{2}}{16}\frac{\nu^{2}}{\mu^{2}}\right|^{1/2}}  \e^{-2\left( S+\frac{\pi}{4}\right) }  .
\end{equation}

Similarly the asymptotic expressions for Linde's tomogram are for $ S<0 $
\begin{align}\label{negativeBi}
\mathcal{W}_{L}(X,\mu,\nu)&\approx   \frac{\mathrm{c}^2}{8\pi^{2}\hbar|\mu|}\frac{(2\hbar\lambda)^{1/3}}{\left|1-\frac{\lambda X}{\mu}-\frac{\lambda^{2}}{16}\frac{\nu^{2}}{\mu^{2}}\right|^{1/2}}
\times \left|   \sin\left(S \right)  \right|^{2} , \\  
\end{align}
and 

\begin{equation}\label{positiveBi}
\mathcal{W}_{L}(X,\mu,\nu)\approx   \frac{\mathrm{c}^2}{8\pi^{2}\hbar|\mu|}\frac{(2\hbar\lambda)^{1/3}}{\left|1-\frac{\lambda X}{\mu}-\frac{\lambda^{2}}{16}\frac{\nu^{2}}{\mu^{2}}\right|^{1/2}}  \e^{2\left( S+\frac{\pi}{4}\right) } .
\end{equation}
for  $ S>0 $

Finally  asymptotic form of Vilenkin's tomogram for $ S<0 $  is
\begin{equation}\label{possibletomogram}
\mathcal{W}_{V}(X,\mu,\nu)\approx   \frac{\mathrm{b}^2}{8\pi^{2}\hbar|\mu|}\frac{(2\hbar\lambda)^{1/3}}{\left|1-\frac{\lambda X}{\mu}-\frac{\lambda^{2}}{16}\frac{\nu^{2}}{\mu^{2}}\right|^{1/2}}\left| \e^{-iS}\right|^{2}.
\end{equation}
and for $ S>0 $
\begin{align}\label{positiveVi}
\mathcal{W}_{V}(X,\mu,\nu)&\approx   \frac{\mathrm{c}^2}{8\pi^{2}\hbar|\mu|}\frac{(2\hbar\lambda)^{1/3}}{\left|1-\frac{\lambda X}{\mu}-\frac{\lambda^{2}}{16}\frac{\nu^{2}}{\mu^{2}}\right|^{1/2}}\\ &\times\left(  \e^{2\left(S+\frac{\pi}{4} \right) }+\frac{\e^{-2\left(S+\frac{\pi}{4} \right) }}{4}\right) 
\end{align}
We fix the normalization coefficients a, b and c  imposing that the quantum and classical tomograms have the same coefficient,  
\begin{equation}\label{Acoefficient}
\mathrm{a}=\frac{\mathrm{b}}{2}=\mathrm{c}=\frac{2^{5/6}\pi\hbar^{1/3}}{\lambda^{1/6}}.
\end{equation}

Therefore we  draw the following conclusions.  First we  notice that in the limit  $  (2\hbar\lambda)^{2/3}\to 0 $ for   $ S>0 $  the tomograms  go either to zero or  to infinity  and they do not match the classical model (\ref{classicaluniversedetomogram}).
For $ S<0 $,  both the Hartle-Hawking  and Linde tomograms do not have a limit  due to the presence of the   the cosine square  and   sine square factors  which go to infinite oscillations. On the other side Vilenkin tomogram  converges to   (\ref{classicaluniversedetomogram}) due to the fact that it contains only the expanding model.

However when we take the limit  $ \lambda\to \lambda_{today} $ with $ \hbar $ constant, we see that   the  extremely small value of  $ \lambda_{today}$  which is of the order of $   10^{-122} \lambda_{Planck} $  justifies  taking the expressions  (\ref{negativeAi}), (\ref{negativeBi}) and (\ref{possibletomogram} ) of the tomograms.  While  Vilenkin's tomogram   becomes again   the classical tomogram,  the    Hartle-Hawking and Linde tomograms have now a limit and become  good candidates to represent a (pseudo-)classical universe.  In other word the decay of the cosmological constant to the present value is responsible for the  smooth transition of the universe  to a  the classical regime.  The Hartle-Hawking and Linde tomograms  present an interference pattern  which  should be  observable, at least in principle, with a large scale survey of the universe.

\section{General relations between the asymptotic solutions and the   classical models }\label{swkb}
The results found in the previous section suggest that to study the relation between quantum and classical cosmological states with any  potential  $ V(q) $ , it is sufficient to construct the quantum tomograms from  the asymptotic solutions of  the Wheeler-DeWitt equation,  whereas the classical tomograms are derived  by introducing a distribution function according to the criterion introduced  in sect.\ref{scosmologicaltomography}.

To  determine the form of the classical tomograms  we consider the  Hamiltonian  
\begin{equation}\label{Hamiltoniangeneralizzata}
\mathcal{H}=\frac{1}{2}\left( -4p^{2}+ V(q)\right) 
\end{equation}
and associate to it the phase space distribution,
\begin{equation}\label{psdistribuzioneconmateria}
g(q,p)=\delta\left( -4p^{2}+V(q)\right) 
\end{equation}
where $ V(q) $ is a generic potential.
We split  $ g(q,p) $ in the following way 
\begin{equation}\label{psdistribuzioneconmateriasplit}
g(q,p)=\frac{\delta\left( p-\frac{\sqrt{V(q)}}{2}\right) }{|\sqrt{V(q)}|}+\frac{\delta\left( p+\frac{\sqrt{V(q)}}{2}\right) }{|\sqrt{V(q)}|}
\end{equation}
where we used  the rule that for any $ F (x) $,  
\begin{equation}\label{delta di F}
\delta(F(x))=\frac{\delta(x-x_{0})}{|F'(x_{0})|}
\end{equation}
if  $ x_{0} $ is the only  solution of equation $ F(x)=0 $ and 
\begin{equation}\label{soluzionimultipledeltadirace}
\delta(F(x))=\Sigma_{i}\frac{\delta(x-x_{i})}{|F'(x_{i})|}.
\end{equation}
if  $ F(x)=0 $  has more solutions $ x_{i} $. 
The tomogram is then given by 
\begin{align}\label{general classical tomogram}
\nonumber \mathcal{W}(X,\mu,\nu)&=\bigintss  \left[ \frac{\delta\left( p-\frac{\sqrt{V(q)}}{2}\right) }{|\sqrt{V(q)}|}+\frac{\delta\left( p+\frac{\sqrt{V(q)}}{2}\right) }{|\sqrt{V(q)}|}\right]  \delta(X-\mu q-\nu p) dq\, dp\\
\nonumber&=\frac{1}{|\nu|}\bigintss  \left[ \frac{\delta\left( \frac{X-\mu q}{\nu}-\frac{\sqrt{V(q)}}{2}\right) }{|\sqrt{V(q)}|}+\frac{\delta\left( \frac{X-\mu q}{\nu}+\frac{\sqrt{V(q)}}{2}\right) }{|\sqrt{V(q)}|}\right]   dq\\
&=\frac{2}{|\nu|}\left( \frac{1}{\left| -\frac{2\mu \sqrt{V(q_{(1)})}}{\nu}-V'(q_{(1)})\right| }+  \frac{1}{\left| -\frac{2\mu \sqrt{V(q_{(2)})}}{\nu}+V'(q_{(2)})\right| }\right) 
\end{align}
where $ q_{(1)}=q_{(1)}(X,\mu,\nu) $ and $ q_{(2)} =q_{(2)}(X,\mu,\nu)$ are  the solutions of the equations
\begin{equation}\label{q1}
\frac{X-\mu q_{(1)}}{\nu}-\frac{\sqrt{V(q_{(1)})}}{2}=0
\end{equation}
and
\begin{equation}\label{q2}
\frac{X-\mu q_{(2)}}{\nu}+\frac{\sqrt{V(q_{(2)})}}{2}=0
\end{equation}
If eq.\. (\ref{q1})  has $ m  $ solutions $ q_{1}^{1}\dots q_{1}^{m} $  and eq.\,(\ref{q2}) has $ n $  solutions $q_{2}^{1}\dots q_{2}^{n} $  the tomogram becomes
\begin{equation}\label{more solutions}
\mathcal{W}(X,\mu,\nu)=\frac{2}{|\nu|}\left(\sum_{i=1}^{m} \frac{1}{\left| -\frac{2\mu \sqrt{V(q_{(1)}^{i})}}{\nu}-V'(q_{(1)}^{i})\right| }+ \sum_{j=1}^{n} \frac{1}{\left| -\frac{2\mu \sqrt{V(q_{(2)}^{j})}}{\nu}+V'(q_{(2)}^{j})\right| }\right) 
\end{equation}
Now let us look for the necessary steps to construct   the quantum tomograms  from the asymptotic  solutions of  the Wheeler-DeWitt equation,
\begin{equation}\label{diffeq}
4\hbar^{2}\frac{d^2\psi(q)}{dq^{2}}-V(q)\psi(q)=0\, .
\end{equation}
In the limit $  \hbar\to 0 $  by applying the WKB method  we find the solutions of  eq. (\ref{diffeq}), whose leading terms 

for $ V(x)>0 $  the solution is 
\begin{equation}\label{soluzvmaggiore}
\psi(q)\approx \frac{A}{|V(q)|^{\frac{1}{4}}}   e^{\pm\int^{x}\sqrt{V(y)}dy},
\end{equation}
and for $ V(q)<0 $
\begin{equation}\label{soluzvminore}
\psi(q)\approx \frac{A}{|V(q)|^{\frac{1}{4}}}  e^{\pm   i\int^{q}\frac{\sqrt{V(y)}}{\varepsilon}dy}.
\end{equation}
where $ A $ is a normalization constant.

We calculate the partial Fourier transforms of the  oscillating functions (\ref{soluzvminore}) because it is where  we can apply the Hartle criterion\cite{Hartle:1986gn} according to which the correlation between the variables can be found only where the wave function is strongly peaked, while the functions (\ref{soluzvmaggiore}) either grow to infinite or go  to zero rapidly.  
This criterion gives a necessary condition  to find a classical solution. Indeed we  calculate the tomogram by applying   the Laplace method to  the integral
\begin{equation}\label{psiasymtotica}
\psi(X,\mu,\nu)=
\int \frac{A}{|V(q)|^{\frac{1}{4}}}  e^{\pm \frac{1}{2\hbar}  i\int^{q}\frac{\sqrt{V(y)}}{\varepsilon}dy +i\frac{\mu q^{2}}{2\hbar \nu}-i\frac{Xq}{\nu}}dq
\end{equation}
we  develop  the exponent to the second order to find 
\begin{align}\label{psiasymptoticlaplace}
\nonumber \psi(X,\mu,\nu)=
\int \frac{A}{|V(q_{0})|^{\frac{1}{4}}}  \exp\left[ {\pm \frac{1}{2\hbar}  i\int^{q_{0}}\sqrt{V(y)}dy +i\frac{\mu q_{0}^{2}}{2\hbar \nu}-i\frac{Xq_{0}}{\hbar\nu}}\right. \\  +\frac{i}{\hbar}\left( \pm \sqrt{V(q_{0})} +\frac{\mu}{\nu}q_{0}- \frac{X}{\nu}\right)(q-q_{0})  \left.+\frac{i}{2\hbar}\left(\pm \frac{1}{2}\frac{V'(q_{0})}{\sqrt{V(q_{0})}}+\frac{\mu}{\nu}\right)(q-q_{0})^{2} \right]dq
\end{align}
where $ q_{0} $ is a stationary point so  that  the first derivative must be zero. This  corresponds precisely to solve  eqs. (\ref{q1}) and (\ref{q2}). 

The  condition on the first derivative is  the correlation condition between momenta and coordinates because from (\ref{X})
\begin{equation}
p= \frac{X-\mu q}{\nu}  
\end{equation}
and  $ \sqrt{V(q)} $ is the derivative of the phase term of the wave function (\ref{soluzvminore}),  
\begin{equation}\label{correlation}
p=\pm \sqrt{V(q)}\,.
\end{equation}
which is equivalent to find the peaks of the function $ \psi(q)\exp(i\tfrac{\mu q^{2}}{2\nu}-i\tfrac{Xq}{\nu})$.

Furthermore   (\ref{correlation})  makes us recognize that the negative  and positive exponents in (\ref{soluzvminore}) represent respectively  the expanding and contracting modes of the universe due to the relation  (\ref{pmomentum}).

After calculating  the Gaussian integral we obtain,
\begin{equation}\label{psi1}
\psi_{1}(X,\mu,\nu)=\left| \sqrt{\frac{4\pi\hbar}{V'(q_{(1)})+2\frac{\mu}{\nu}\sqrt{V(q_{((1))})}}}\right| 
\mathrm{ e}^{\left( \frac{i}{\hbar}\int ^{q_{(1)}} V(y)dy+\frac{i\mu q_{(1)}^{2}}{\hbar\nu}-\frac{iXq_{(1)}}{\hbar\nu}\right)}
\end{equation}
and
\begin{equation}\label{psi2}
\psi_{2}(X,\mu,\nu)=\left| \sqrt{\frac{4\pi\hbar}{-V'(q_{(2)})+2\frac{\mu}{\nu}\sqrt{V(q_{(2)})}}}\right| 
\mathrm{e}^{\left( -\frac{i}{\hbar}\int ^{q_{(2)}} V(y)dy+\frac{i\mu q_{(2)}^{2}}{\hbar\nu}-\frac{iXq_{(2)}}{\hbar\nu}\right)}\, .
\end{equation}
The general solution is 
\begin{equation}\label{psitot}
\psi(X,\mu,\nu)=c_{1} \psi_{1}(X,\mu,\nu)+ c_{2}\psi_{2}(X,\mu,\nu)
\end{equation}
and the tomogram is given by 
\begin{equation}\label{general tomogram}
\mathcal{W}(X,\mu,\nu)=\frac{1}{2\pi\hbar|\nu|}\left|\psi \right|^{2}=\frac{2}{|\nu|}\left[c_{1}^{2}\left|\psi_{1} \right|^{2}+c_{2}^{2}\left|\psi_{2}\right|^{2} +2c_{1}c_{2}\psi_{1}\psi_{2}\right] 
\end{equation}
Among    all the possible choices of $ c_{1} $ and $ c_{2} $,  we find the tomograms related to the    Hartle-Hawking, Vilenkin and Linde models  respectively when  $ c_ {1} = c_ {2} $,  $ c_{2}\neq 0 $ and $ c_{1}=0 $,  and $ c_ {1} =- c_ {2} $.  

In general all  these tomograms, which are now classical, as the left hand side of  the uncertainty principle (\ref{tomographic uncertainty relation}) goes to zero, present  interference  terms as for the Hartle-Hawking and Linde tomograms in the de Sitter universes.  

Only the set of tomograms obtained  from the  combination
\begin{equation}\label{quantum to classical tomogram}
\psi(X,\mu,\nu)= \psi_{1}(X,\mu,\nu)+ i \psi_{2}(X,\mu,\nu)\, ,
\end{equation} 
coincides with  (\ref{general classical tomogram}). But when  eqs. (\ref{q1}) and (\ref{q2}) have multiple solutions, we have that 
\begin{equation}\label{psi1multiplesoluzioni}
\psi_{1}(X,\mu,\nu)=\sum_{i=1}^{m}\left| \sqrt{\frac{4\pi\hbar}{V'(q^{i}_{(1)})+2\frac{\mu}{\nu}\sqrt{V(q^{i}_{((1))})}}}\right| 
\mathrm{ e}^{\left( \frac{i}{\hbar}\int ^{q^{i}_{(1)}} V(y)dy+\frac{i\mu {q^{i}}_{(1)}^{2}}{\hbar\nu}-\frac{iXq^{i}_{(1)}}{\hbar\nu}\right)}
\end{equation}
and
\begin{equation}\label{psi2multiplesoluzioni}
\psi_{2}(X,\mu,\nu)=\sum_{j=1}^{n}\left| \sqrt{\frac{4\pi\hbar}{-V'(q^{j}_{(2)})+2\frac{\mu}{\nu}\sqrt{V(q^{j}_{(2)})}}}\right| 
\mathrm{e}^{\left( -\frac{i}{\hbar}\int ^{q^{j}_{(2)}} V(y)dy+\frac{i\mu {q^{j}}_{(2)}^{2}}{\hbar\nu}-\frac{iXq^{j}_{(2)}}{\hbar\nu}\right)}\, .
\end{equation}
and the tomograms (\ref{general tomogram}) obtained from (\ref{psi1multiplesoluzioni}) and (\ref{psi2multiplesoluzioni}) never  converge to   (\ref{more solutions}) due to the unavoidable presence of the mixed products. 

\section{Models with the cosmological constant}\label{smodelswithcosmologicalconstant}
So far we have considered general models with any  potential and we found the conditions of compatibility between quantum and classical tomograms. However, no obvious criteria appear for the transition to a classical universe. In this section  we want to understand if the decay  of the  cosmological constant can responsible for this transition as in the case of de Sitter  universes.  Therefore we will consider the potentials that contain the cosmological constant explicitly

\begin{equation}\label{diffeqlambda1}
4\hbar^{2}\frac{d^2\psi(q)}{dq^{2}}+(\lambda q-1 - \varepsilon f(q)) \psi(q)=0\, ,
\end{equation}
and we can write the Wheeler-DeWitt equation as a modification of   eq.(\ref{WheelerDeWitt}).

We can face this problem in three different ways. If $ f(q) $ is a perturbative term we can solve the equation eq. (\ref{diffeqlambda1}) by a regular perturbation method\cite{Zwillinger}, the second is to solve it by taking  an uniform asymptotic expansion of  solutions \cite{abramowitz}, the third one by applying the method illustrated in the previous section.

In the first case,with the change of variables  (\ref{changeofvariable1}), we have
\begin{equation}\label{Airyequationperturbed}
\frac{d^{2}\psi(\xi)}{d \xi^{2}}-\left( \xi+\varepsilon f\left( \dfrac{1-(2\hbar\lambda)^{2/3}\xi}{\lambda}\right) \right) \psi(\xi)=0,  
\end{equation}
We consider the expansion 
\begin{equation}\label{expansion}
\psi(\xi)=\psi_{0}(\xi)+\varepsilon\psi_{1}(\xi)+\dots
\end{equation}
and we obtain the equations
\begin{equation}\label{epsilonugualezero}
\frac{d^{2}\psi_{0}(\xi)}{d \xi^{2}}- \xi \psi_{0}(\xi)=0,  
\end{equation}
\begin{equation}\label{epsilonuguale1}
\frac{d^{2}\psi_{1}(\xi)}{d \xi^{2}}-\xi\psi_{1}=f\left( \frac{1-(2\hbar\lambda)^{2/3}\xi}{\lambda}\right) \psi_{0}(\xi),  
\end{equation}
this second equation is an inhomogeneous Airy equation.
Let's take  for sake of brevity $ \psi_{0}(\xi)=\Ai(\xi) $, the perturbed solution is
\begin{align}\label{soluzioneperturbata}
\nonumber \psi(\xi)  &=\left( 1+\varepsilon \left( c_{1}+\pi\int\Ai(\xi)\Bi(\xi)f\left( \frac{1-\xi}{\lambda}\right)d\xi\right)  \right) \Ai(\xi) \\
& +\varepsilon \left( c_{2} -\pi\int\Ai^{2}(\xi)f\left( \dfrac{1-(2\hbar\lambda)^{2/3}\xi}{\lambda}\right)d\xi\right)  \Bi(\xi)
\end{align}

Where $ c_{1} $ and $ c_{2} $  are arbitrary integration constants. The solutions are proportional to the unperturbed de Sitter solutions,  so from eq. (\ref{nunonzero}) we derive the perturbed de Sitter tomograms provided the integrals of the inhomogeneous solution are well-behaved.

Very similar results come from the application of the uniform asymptotic expansion of solutions\cite{abramowitz},  where if $ \xi $ varies on a bounded interval $ a\leq\xi\leq b $  and where for each fixed  $ (2\hbar\lambda)^{-2/3} $,  $ f\left( \frac{1-(2\hbar\lambda)^{2/3}\xi}{\lambda}\right)  $ is continuous in$ \xi $ for  $ a\leq\xi\leq b $ there are solutions $ \psi_{a} (\xi)$ and $ \psi_{b}(\xi) $  such that uniformly on $ \xi $ for $ a\leq\xi\leq  0$
\begin{align}
\nonumber\psi_{a}(\xi)& = \Ai(-\xi)(1-O( (2\hbar\lambda)^{-2/3}))\ \ \ \ \ \ \ (2\hbar\lambda)^{-2/3}\to 0 \\
\psi_{b}(\xi)& = \Bi(-\xi)(1-O( (2\hbar\lambda)^{-2/3}))
\end{align}
and 
uniformly on $ \xi $ for $ 0\leq\xi\leq  b$
\begin{align}
\nonumber \psi_{a}(\xi)& = \Ai(-\xi)(1-O( (2\hbar\lambda)^{-2/3})) +\Bi(-\xi)O((2\hbar\lambda)^{-2/3}) \ \ \ \ \ \ \ (2\hbar\lambda)^{-2/3}\to 0\\
\psi_{b}(\xi)& = \Bi(-\xi)(1-O( (2\hbar\lambda)^{-2/3})) +\Ai(-\xi)O(2\hbar\lambda)^{-2/3} 
\end{align}
Finally to apply the WKB method, we   take  the change of variables $ y=1-\lambda q $ and the Wheeler-DeWitt becomes
\begin{equation}\label{diffeqlambda2}
4\hbar^{2}\lambda^{2}\frac{d^2\psi(y)}{dy^{2}}-\left( \xi -f\left( \frac{1-y}{\lambda}\right)  \right) \psi(y)=0\, .
\end{equation}
And we consider the oscillating asymptotic solutions,
\begin{equation}\label{asymptotic}
\psi(y)=\frac{1}{\sqrt[4]{\left|\xi+g\left( \frac{1-y}{\lambda}\right)  \right| }}
\mathrm{e}^{\pm\frac{i}{2\hbar\lambda}\int{\sqrt{y+g\left( \frac{1-y}{\lambda}\right)  }}}dy
\end{equation}
we then proceed calculating  the fractional Fourier transform and we recover all the results of sect. \ref{swkb}. 

We conclude that also in this case  the decay of the cosmological constant plays a  crucial role in the transition to a classical universe.

\section{Conclusions}\label{sConclusions}

Symplectic tomography offers us the possibility to express quantum and classical  cosmological states in an unified formalism giving us the possibility to compare them and  to determine when  a quantum to classical transition is possible.
This approach allowed us to study in general the de Sitter models. The results of  \cite{Stornaiolo:2018lvp} and in sect.\ref{scosmologicaltomography}  have been recalled and the  relation of the various proposals of  initial conditions of the universe with the classical tomogram (\ref{classicaldeSitterBoltzmann}) were examined. We found that only the Vilenkin tomogram converges to a classical universe, while the Hartle and Hawking as well as the Linde's tomograms do not have a limit. However when we consider the decay of the cosmological constant $ \lambda\to \lambda_{today} $  we can take as well the  tomograms in their asymptotic form and  the universe enters to a classical state.   Because this limit  is to a  finite value  we are led to  consider also the Hartle and Hawking and Linde as  "classical" models with interference terms which are  peculiar and unexpected properties if  observed.  

These interference terms are a general characteristic in  most of the quantum tomograms and their classical limits  as it came out in  sect.\ref{swkb}. The quantum tomograms   converge to the  classical tomogram (\ref{general classical tomogram}) only when they satisfy the condition (\ref{quantum to classical tomogram}) and in some particular cases.

Here by classical we mean tomograms obtained  from the classical Einstein equations,  but we could also   consider as \emph{classical}  all the  tomograms obtained asymptotically as they violate the uncertainty principle.  They may  equally describe the present universe correctly.   Finally  in sect.\ref{smodelswithcosmologicalconstant}  we  established the criteria by which the cosmological constant can induce the transition to a classical universe.

Although we have found general  results in this paper in the tomographic analysis of the quantum-to-classical transiton,   we want to highlight that there are some points that still need to be explored in the future work.

For example, the fractional Fourier  transforms we have used so far are all defined in an interval ranging from $ - \infty $ to $ + \infty $ even if variable $ q $ is explicitly positive. Therefore it may seem incorrect to use it in general. There are different way around this problem, for example we may consider only wave functions with a compact support  , or we  can  an infinite barrier at $ q=0 $  in the potential and finally instead of using  the fractional Fourier transform  we can introduce the fractional Hankel  transform    (see for example\cite{Hankel2} and \cite{Hankel1}) where the integration range runs  from $ 0 $ to  $ + \infty $.

As we have seen, the decay of the cosmological constant could explain two crucial points,  why the quantum universe evolved  to a classical state and the smallness of the cosmological constant\cite{Weinberg:1988cp}.  This  phenomenon is not new in cosmology  as it is used in many models to  explain  the exit from inflation. But now  this problem has to be addressed in  a quantum universe, even by considering  models with scalars fields or   other ways. Some examples are given  in  \cite{Moffat:2014zwa}   \cite{Mikovic:2014opa}  \cite{Polyakov:2009nq}. 

Further we can  broaden the results found to more general  forms of  dark energy. These can  be  treated in terms of  extended theories $ F (R) $. In this case it might be interesting to see if  these theories lead to a  self-consistent way to describe the quantum to classical  transition.

So far we have discussed the mathematical implications of the tomographic representation.  We recall here  that the  important property of tomograms is that they  observable  and therefore it is presumable that we can build a model of the universe in tomographic terms based on the observable quantities such as such as the Hubble constant $H_{0}$, the deceleration parameter $ q_{0}$. and the density parameters $ \Omega_{k} $, $ \Omega_{\Lambda} $ and $ \Omega_{m} $.

In the de Sitter models the argument of the tomogram can be interpreted by writing explicitly   $ X=\mu q + \nu p $ in the argument of the tomogram and using the equations we find 
\begin{align}\label{parametri cosmografici}
\nonumber 1- \lambda \frac{X}{\mu}  -\frac{\lambda^{2}}{16}\frac{\nu^{2}}{\mu^{2}}& =    1- \lambda q -\frac{\lambda \nu}{\mu}p+\frac{\lambda^{2}}{16}  \frac{\nu^{2}}{\mu^{2}}\\
&=1-\lambda q\left( 1-\frac{1}{4}\frac{\nu}{\mu}\frac{\dot{q}}{q}
+\frac{1}{24}\frac{\nu^{2}}{\mu^{2}}\frac{\ddot{q}}{q}\right)  
\end{align}
where we have used  $ p=-\frac{\dot{q}}{4} $ and 
$ \ddot{q}=\frac{2}{3}\lambda  $ according  and to (\ref{equationsinloukoform2}) so we see  that these terms are related to the Hubble constant and the deceleration parameter and suggest that  $ \nu/\mu \propto  \Delta \tau $.  In the other tomograms we expect more complicated relations between  the variables ($ X, \mu, \nu $) and the cosmological observables  obtained after solving the equations (\ref{q1}) and (\ref{q2})  . 

Tomograms are probability functions, so we adopt the point of view that  they  describe  the statistics of  a universe  made by a patchwork of regions with the same  homogeneous metric, but  where the parameters that  characterize them can vary. This happens, for example, when  cosmological perturbations are involved.  Then the  statistical description  of the observations  will give us more complete information on the nature of the potential  $ V(q) $ and therefore a more precise knowledge of the initial state of the universe.

\section*{Acknowledgments}

This work was started and completed during the Covid-19 quarantene when all our personal relations where limited to our family. I spent all this time with my wife Rita Maria  and her son Federico  and I am very glad to thank them for  all the support they give to me while preparing and writing this   paper.

\end{document}